\documentstyle[preprint,aps,epsfig]{revtex}

 \newcommand{\bq}{\begin{equation}}
 \newcommand{\eq}{\end{equation}}
 \newcommand{\bqn}{\begin{eqnarray}}
 \newcommand{\eqn}{\end{eqnarray}}
 \newcommand{\nb}{\nonumber}
 \newcommand{\lb}{\label}

\begin{document}

\title{\LARGE Asymptotics of Solutions of a Perfect Fluid  Coupled with a Cosmological
Constant in Four-Dimensional Spacetime with Toroidal Symmetry}  
\author{Gregory A. Benesh and Anzhong Wang}
\address{Department of Physics, Baylor University, 101 Bagby Avenue, Waco, Texas
76706}
\date{\today}

\maketitle

\begin{abstract}
Asymptotics of solutions of a perfect fluid  when coupled with a cosmological
constant in four-dimensional spacetime with toroidal symmetry are studied. 
In particular, it is found that the problem of self-similar solutions of the 
first kind for a fluid with  the equation of state, $p = k \rho$, can be 
reduced to solving a master equation of the form,
$$
2 F(q, k)\frac{q''(\xi)}{q'(\xi)}   - G(q,k)  q'(\xi) = \frac{4}{\xi}.
$$
For $k = 0$ and $k = -1/3$ the general solutions are obtained and their main 
local and global properties are studied in detail.

\end{abstract}

\vspace{4.cm}



\section{Introduction}
\renewcommand{\theequation}{1.\arabic{equation}}
 \setcounter{equation}{0}

Recently, self-similar solutions of the Einstein field equations have attracted 
much attention, not only because they   can be studied analytically
through simplification of the problem, but also because of
their relevance in  astrophysics  \cite{Car} and critical phenomena in  gravitational 
collapse \cite{Chop93,Wang01,Gun03}. In the latter case, as
in statistical mechanics \cite{Bar79,Go92}, the self-similar solutions represent 
the asymptotics of critical collapse in the intermediate regions.   

Recently, self-similar solutions of a scalar field  were studied and  
all such solutions were found analytically \cite{W04}. It was shown that some of the 
solutions can be interpreted as representing gravitational collapse of the scalar field. 
During the collapse, trapped surfaces   never develop, and as a result, no black hole is 
formed. This is consistent with the general theorem presented in \cite{Wang03}. Although the 
collapse always ends in spacetime singularities, it was found that these singularities 
are spacelike and not naked \cite{Joshi}.

In this paper, we shall study self-similar solutions of a perfect fluid with toroidal symmetry
and the equation of state, $p = k \rho$, where $k$ is a constant, and $p$ and $\rho$ denote,
respectively, the pressure and energy density of the fluid. In particular, in
Sec. II we  write  the corresponding Einstein field equations and define 
apparent horizons and black holes in such spacetimes, while
in Sec. III we present the dimensional analysis of the problem. In Sec. IV, following
Barenblatt \cite{Bar79}, we study the self-similar solutions of the first kind, and
find general solutions for $k = 0$ and $k =  -1/3$. For other cases the problem reduces
 to solving  a master equation of the form,
\bq
\lb{1.1}
2 F(q, k)\frac{q''}{q'}   - G(q, k)  q' = \frac{4}{\xi}.
\eq 
In Sec. V, we study the local and global properties of the solutions found in Sec. IV,
while in Sec. VI, we present our main conclusions.

\section{The Einstein-Fluid Equations }

\renewcommand{\theequation}{2.\arabic{equation}}
 \setcounter{equation}{0}

The general metric for a four-dimensional spacetime with toroidal symmetry 
can be cast in the form,
 \bq
 \lb{2.1}
 ds^2 = g_{ab}\left(x^{0}, x^{1}\right)dx^{a} dx^{b}
 - {L_{0}}^{2}e^{-\mu\left(x^{0}, x^{1}\right)}\left(d\theta^{2}   
 + d\varphi^{2}\right),\;\; (a, b = 0, 1),
 \eq
where   $x^{a}\in (-\infty,\infty)$ and  all $x^{a}$ have  the dimension of length, $L$.  
The coordinates $x^{A} = \{\theta,\; \varphi\},\; (A = 2, 3)$ are dimensionless with the 
surfaces $x^{A} = 0$ and $ x^{A} = 2\pi$ being identified. The constant $L_{0}$ has the 
dimension of length. The spacetimes usually have three Killing vectors, given, 
respectively, by
 \bq
 \lb{2.2}
 \xi_{(1)} = \theta  \frac{\partial}{\partial \varphi}
 - \varphi \frac{\partial}{\partial \theta }, \;\;\;
 \xi_{(2)} = \frac{\partial}{\partial \theta },\;\;\;
 \xi_{(3)} = \frac{\partial}{\partial \varphi}.
 \eq
  Clearly, the metric is
invariant under the coordinate transformation,
 \bq
 \lb{2.3}
 x^{0} = x^{0}\left({x'}^{0}, {x'}^{1}\right),\;\;\;\;
 x^{1} = x^{1}\left({x'}^{0}, {x'}^{1}\right).
 \eq
The energy-momentum tensor (EMT) for a perfect fluid, , 
takes the form
 \bq
 \lb{2.4}
 T_{\mu\nu} = \left(\rho + p\right)u_{\mu}u_{\nu} 
 - p g_{\mu\nu},
 \eq
where $u_{\mu},\; \rho$ and $p$ denote, respectively, the four-velocity, 
energy density and pressure of the fluid.
Using  gauge freedom (\ref{2.3}) we  choose the coordinates such 
that 
\bq
\lb{2.5}
g_{01} \left(x^{0}, x^{1}\right) =0, \;\;\; 
u_{\mu} = \left(g_{00}\right)^{1/2} \delta^{0}_{\mu}.
\eq
Denoting    $x^{0}$ and $ x^{1}$ by $t$ and $z$, respectively, 
we find that the metric can be written as
 \bq
 \lb{2.6}
 ds^2 = e^{\lambda(t,z)}dt^{2} - e^{\nu(t,z)}dz^{2} 
 - {L_{0}}^{2}e^{-\mu(t,z)}\left(d\theta^{2} + d\varphi^{2}\right).
 \eq
In this note we    consider asymptotics of solutions of a perfect fluid. 
Using the remaining (trivial) coordinate transformation $ t' = t + t_{0}$,
where $t_{0}$ is a constant, we     assume that a spacetime singularity
always starts to form at $t = 0$ (if there is any).  Then, it can be shown 
that the Einstein-fluid equations,
\bq
\lb{2.8}
  R_{\mu\nu} - \frac{1}{2}g_{\mu\nu}R + \Lambda g_{\mu\nu}
 = \left(\rho + p\right)u_{\mu}u_{\nu} 
 - p g_{\mu\nu},
\eq
 can be cast in the form,
\bqn
 \lb{2.9a}
& & 2\mu_{,tz} - \mu_{,t}\left(\mu_{,z} + \lambda_{,z}\right) 
 - \mu_{,z}\nu_{,t} = 0,\\
\lb{2.9b}
& &  e^{\lambda -\nu}\left\{2\left(\lambda_{,zz} - \mu_{,zz}\right)
+ \mu_{,z}\left(\mu_{,z} + \lambda_{,z} + \nu_{,z}\right)
+ \lambda_{,z}\left(\lambda_{,z} - \nu_{,z}\right)
\right\} \nb\\
& & -   2\left(\nu_{,tt} + \mu_{,tt}\right)
- \mu_{,t}\left(2\mu_{,t} +   \nu_{,t} + \lambda_{,t}\right)
+ \nu_{,t}\left(\nu_{,t} - \lambda_{,t} \right) = 0,\\
\lb{2.10a}
& & \rho = \frac{1}{4}\left\{e^{- \nu}\left[4\mu_{,zz}
- \mu_{,z}\left(3\mu_{,z} + 2\nu_{,z}\right)\right]
+ e^{- \lambda} \mu_{,t}\left(\mu_{,t} - 2\nu_{,t}\right)
\right\} + \Lambda,\\
\lb{2.10b}
& & p = \frac{1}{4}\left\{e^{ - \lambda}\left[4\mu_{,tt}
- \mu_{,t}\left(3\mu_{,t} + 2\lambda_{,t}\right)\right]
+ e^{ - \nu}\mu_{,z}\left(\mu_{,z} - 2\lambda_{,z}\right)
\right\} -  \Lambda,
\eqn
where $\Lambda$ denotes the cosmological constant, which can be parameterized as
\bq
\lb{2.11}
\Lambda = \pm \frac{1}{l^{2}}.
\eq
 
To study the formation of apparent horizons and black holes
in the spacetimes described by the metric (\ref{2.6}), we    first give
the definition of such notions. Following \cite{Wang03,Wang} 
let us first introduce two null coordinates $u$ and $v$ via the relations
  \bq
  \lb{B.1}
  du = J(t,z)\left(e^{\lambda/2} dt - e^{\nu/2}dz\right),\;\;\;
  dv = K(t,z)\left(e^{\lambda/2} dt + e^{\nu/2}dz\right),
  \eq
 where $J(t,z)$ and $K(t,z)$ satisfy the integrability conditions,
\bq
\lb{B.1aa}
\frac{\partial^{2} u}{\partial t\partial z} = 
\frac{\partial^{2} u}{\partial z\partial t},\;\;\;\;
\frac{\partial^{2} v}{\partial t\partial z} = 
\frac{\partial^{2} v}{\partial z\partial t}.
\eq
Without loss of generality, we  assume that they are all strictly positive,
 \bq
 \lb{B.1a}
 J(t,z) > 0,\;\;\;\; K(t,z) > 0.
 \eq
Then, it is easy to show that, in terms of $u$ and $v$,
 the metric (\ref{2.6}) takes the form
\bq
\lb{B.2}
ds^{2} =  2 e^{2\sigma(u, v)} dudv -
{\cal R}^{2}(u,v) \left(d\theta^{2} + d\varphi^{2}\right),
\eq
 where
\bq
\lb{B.3}
\sigma(u,v) = - \frac{1}{2}\ln\left(2JK\right),\;\;\;
{\cal R}(u,v) = L_{0}e^{-\mu/2}.
\eq
It should be noted that the metric (\ref{B.2}) is invariant under the transformations
\bq
\lb{B.3aa}
u = u(\bar{u}),\;\;\;
v = v(\bar{v}).
\eq
Using this gauge freedom, we  assume that the metric has no coordinate singularities
in the coordinates, $u, \; v$ and $x^{A}$.  This, in particular,  implies that $\sigma$ is 
finite except at some points or on some surfaces where the spacetime is singular.

Introducing two null vectors, $l_{\mu}$ and $n_{\mu}$, by, 
\bq
\lb{B.6}
l_{\lambda} \equiv \frac{\partial u}{\partial x^{\lambda}} =
\delta^{u}_{\lambda},\;\;\;
n_{\lambda} \equiv \frac{\partial v}{\partial x^{\lambda}} =
\delta^{v}_{\lambda},
\eq
which  are orthogonal  to the two-surfaces ${\cal S} = \{x^{\lambda}: t, z = Constant\}$,
we find that the expansions of the null rays $u = Constant$ and   $v = Constant$
are given, respectively, by 
\bqn
\lb{B.6c}
\theta_{l} &\equiv& l_{\mu;\nu}g^{\mu\nu} = 2e^{-2\sigma}\frac{{\cal R}_{,v}}{{\cal R}},
 \nb\\
\theta_{n} &\equiv& n_{\mu;\nu}g^{\mu\nu} = 2e^{-2\sigma}\frac{{\cal R}_{,u}}{{\cal R}}.
\eqn
 
{\em Definitions}. A two-surface, ${\cal{S}}$,  of constant $u$ and $v$ (or constant $t$ and $z$)
is {\em trapped, marginally trapped, or untrapped}, according to whether 
$\theta_{l}\theta_{n} > 0$, $\; \theta_{l}\theta_{n} = 0$, or $\theta_{l}\theta_{n} < 0$.
Assuming that on the marginally trapped surfaces  we have 
$\left.\theta_{l}\right|_{{\cal{S}}} = 0$, an {\em apparent horizon} 
is defined as a two-surface 
$H$ foliated by marginally trapped surfaces, on which $\left.\theta_{n}\right|_{H} \not=0$.  
The apparent horizon is {\em outer, degenerate, or inner}, according to whether
$\left.{\cal{L}}_{u}\theta_{l}\right|_{H} < 0$, $\; \left.{\cal{L}}_{u}\theta_{l}\right|_{H} 
= 0$, or $\left.{\cal{L}}_{u}\theta_{l}\right|_{H} > 0$, {\em future} if 
$\left.\theta_{n}\right|_{H} < 0$,  and  {\em past} if 
$\left.\theta_{n}\right|_{H} > 0$,  where ${\cal{L}}_{v}\;$ (${\cal{L}}_{u}$) denotes the Lie 
derivative along the vector $l^{\mu}\;$ ($n^{\mu}$). 
We define {\em a black hole by the existence of a  future outer or degenerate  apparent horizon}
\cite{Wang03,BHs}.  

It is interesting to note that the above definitions can   also be given in terms of the
vector $ {\cal R}_{,\lambda}$,
\bq
\lb{B.6f}
{\cal R}_{,\lambda} {\cal R}^{,\lambda} 
=2 e^{-2\sigma}  {\cal R}_{,v} {\cal R}_{,u} = \frac{1}{2}{\cal R}^{2}e^{2\sigma}
 \theta_{l}\theta_{n}. 
\eq
Thus, a two-surface, ${\cal{S}}$,  of constant $t$ and $z$  
is {\em trapped, marginally trapped, or untrapped}, according to whether
${\cal{R}}_{,\lambda}$ is timelike, null, or spacelike.  

From Eq.(\ref{B.1}) we find that
  \bqn
  \lb{B.4}
  \frac{\partial t}{\partial u} &=& \frac{1}{2J}e^{-\lambda/2},\;\;\;
  \frac{\partial t}{\partial v} = \frac{1}{2K}e^{-\nu/2},\nb\\
  \frac{\partial z}{\partial u} &=& - \frac{1}{2J}e^{-\nu/2},\;\;\;
  \frac{\partial z}{\partial v} = \frac{1}{2K}e^{-\nu/2},
  \eqn
from which we have  
  \bqn
  \lb{B.5}
  \theta_{l}  &=& 2e^{-2\sigma}\frac{{\cal{R}}_{,v}}{{\cal{R}}}
  = \frac{2K}{{\cal{R}}}\left(e^{-\lambda/2}{\cal{R}}_{,t} 
  + e^{-\nu/2}{\cal{R}}_{,z}\right),\nb\\
  \theta_{n} &=& 2e^{-2\sigma}\frac{{\cal{R}}_{,u}}{{\cal{R}}}
  = \frac{2J}{{\cal{R}}}\left(e^{-\lambda/2}{\cal{R}}_{,t} 
  - e^{-\nu/2}{\cal{R}}_{,z}\right).
  \eqn

\section{Dimensional Analysis}
\renewcommand{\theequation}{3.\arabic{equation}}
 \setcounter{equation}{0}

Since $t$ and $z$ all have dimensions of length, and $\lambda,\; \nu$ and
$\mu$ are dimensionless, we have
\bq
\lb{3.1}
[t] = [z] = L,\;\;\; [\lambda] = [\nu] = [\mu] = 1.
\eq
On the other hand, because the Ricci tensor has the dimension of $L^{-2}$, then from
Eq.(\ref{2.8})  we find that
\bq
\lb{3.2}
\left[\rho\right] = \left[p\right] = L^{-2},
\;\;\;\;  \left[l\right] = L.
\eq
Thus, from $t, \; z$ and $l$ we construct two dimensionless quantities,
\bq
\lb{3.3}
\xi \equiv -  \frac{z}{t}, \;\;\;\; \eta \equiv - \frac{t}{l}.
\eq

Since the metric coefficients are functions of 
$t,\; z$ and $l$, and $\lambda,\; \nu$ and $\mu$ are dimensionless, we must have
\bqn
\lb{3.4a}
& & \lambda\left(t,z,l\right) = \lambda\left(\xi, \eta\right),\;\;\;
\nu\left(t,z,l\right) = \nu\left(\xi, \eta\right),\;\;\;
\mu\left(t,z,l\right) = \mu\left(\xi, \eta\right),\nb\\
\lb{3.4b}
& & \rho\left(t,z,l\right) =\frac{1}{4t^{2}}\Phi\left(\xi, \eta\right),\;\;\;
 p\left(t,z,l\right) =\frac{1}{4t^{2}}\Pi\left(\xi, \eta\right).
\eqn
For any given function $f(\xi, \eta)$,   we have
\bqn
\lb{3.5}
f_{,t} &=& - \frac{1}{t}\left(\xi f_{,\xi} - \eta f_{,\eta}\right),\nb\\
f_{,z} &=& - \frac{1}{t} f_{,\xi},\nb\\
f_{,tt} &=&  \frac{1}{t^{2}}\left(\xi^{2} f_{,\xi\xi} -
2 \eta\xi f_{,\eta\xi} + \eta^{2}f_{,\eta\eta} + 2 \xi f_{,\xi}\right),\nb\\
f_{,tz} &=& \frac{1}{t^{2}}\left(\xi f_{,\xi\xi}- \eta f_{,\eta\xi} +
f_{,\xi}\right),\nb\\
f_{,zz} &=& \frac{1}{t^{2}}f_{,\xi\xi}.
\eqn
Inserting these expressions into Eqs.(\ref{2.9a})-(\ref{2.10b}), we find that
\bqn
\lb{3.6a}
& & 2\left(\xi\mu_{,\xi\xi} - \eta\mu_{,\eta\xi} + \mu_{,\xi}\right)
- \left(\mu_{,\xi} + \lambda_{,\xi}\right)\left(\xi\mu_{,\xi} -
\eta\mu_{,\eta}\right)\nb\\
& & \;\;\;\;\;\; 
- \mu_{,\xi}\left(\xi\nu_{,\xi} -
\eta\nu_{,\eta}\right) = 0,\\
& & 2\left[\xi^{2}\left(\mu_{,\xi\xi} + \nu_{,\xi\xi}\right)
- 2\eta\xi\left(\mu_{,\eta\xi} + \nu_{,\eta\xi}\right)
+ \eta^{2}\left(\mu_{,\eta\eta} + \nu_{,\eta\eta}\right)
+ 2\xi \left(\mu_{,\xi} + \nu_{,\xi}\right)\right]\nb\\
& & -2\left(\xi\mu_{,\xi} -\eta \mu_{,\eta}\right)
\left[\xi\left(\mu_{,\xi} + \lambda_{,\xi}\right)
- \eta\left(\mu_{,\eta} + \lambda_{,\eta}\right)\right]\nb\\
& & + \left[\xi\left(\nu_{,\xi} - \lambda_{,\xi}\right)
- \eta\left(\nu_{,\eta} - \lambda_{,\eta}\right)\right]
 \left[\xi\left(\nu_{,\xi} - \mu_{,\xi}\right)
- \eta\left(\nu_{,\eta} - \mu_{,\eta}\right)\right]\nb\\
& & + e^{\lambda - \nu}\left[2\left(\mu_{,\xi\xi}
- \lambda_{,\xi\xi}\right) - \mu_{,\xi}\left(\lambda_{,\xi} 
+ \nu_{,\xi}\right) -  \lambda_{,\xi}\left(\lambda_{,\xi} 
- \nu_{,\xi}\right)\right] = 0,\\
\lb{3.7a}
 \Phi &=&  e^{-\nu}\left[4\mu_{,\xi\xi} -
\mu_{,\xi}\left(3\mu_{,\xi} + 2\nu_{,\xi}\right)\right]\nb\\
& & + e^{-\lambda}\left(\xi\mu_{,\xi} -\eta \mu_{,\eta}\right)
 \left[\xi\left(\mu_{,\xi} - 2\nu_{,\xi}\right)
- \eta\left(\mu_{,\eta} - 2\nu_{,\eta}\right)\right]
 \pm 4 \eta^{2},\\
\lb{3.7b}
 \Pi &=&  e^{-\lambda}\left\{4\left(\xi^{2}\mu_{,\xi\xi}
- 2\eta\xi\mu_{,\eta\xi} + \eta^{2}\mu_{,\eta\eta}
+ 2\xi \mu_{,\xi}\right)\right.\nb\\
& & \left. - \left(\xi\mu_{,\xi} -\eta \mu_{,\eta}\right)
 \left[\xi\left(3\mu_{,\xi} + 2\lambda_{,\xi}\right)
- \eta\left(3\mu_{,\eta} + 2\lambda_{,\eta}\right)\right]\right\}\nb\\
& & + e^{-\nu}\mu_{,\xi}\left(\mu_{,\xi} - 2\lambda_{,\xi}\right)
\pm 4\eta^{2},
\eqn
where the + (-) sign corresponds to $\Lambda > 0 \; (\Lambda < 0)$.

Note that in the present case  we have four equations, Eqs.(\ref{3.6a})
- (\ref{3.7b}), and  five unknowns: $\lambda,\; \nu,\; \mu,
\; \Phi$ and $\Pi$. Thus, to determine these five functions uniquely, 
we need to have one more equation, which is usually provided by the 
equation of state of the perfect fluid. In this paper, we    consider 
the case 
\bq
\lb{3.8}
p = k \rho,
\eq
where $k$ is a constant. 

\section{Self-Similar Solutions of the First Kind}
\renewcommand{\theequation}{4.\arabic{equation}}
 \setcounter{equation}{0}
 
 In this section, we consider the asymptotic behavior of the previous 
equations as $\eta \rightarrow 0$.  
According to Barenblatt \cite{Bar79} (See also \cite{Go92}), self-similar 
solutions of the first kind are defined as the existence of the limits,
 \bqn
 \lb{4.1}
\lambda_{0}(\xi) &=& \lim_{\eta \rightarrow 0}{\lambda(\xi, \eta)},\;\;\;
\nu_{0}(\xi) = \lim_{\eta \rightarrow 0}{\nu(\xi, \eta)},\nb\\
\mu_{0}(\xi) &=& \lim_{\eta \rightarrow 0}{\mu(\xi, \eta)},\;\;\;
\Phi_{0}(\xi) = \lim_{\eta \rightarrow 0}{\Phi(\xi, \eta, p)},\nb\\
\Pi_{0}(\xi) &=& \lim_{\eta \rightarrow 0} {\Pi(\xi,\eta)}.
\eqn
Substituting these expressions into Eqs.(\ref{3.6a})-(\ref{3.7b}) and
Eq.(\ref{3.8}), we find that
\bqn
\lb{4.2a}
& & 2\left(\xi\mu'\right)' - \xi\mu'\left(\lambda'
+ \nu' + \mu'\right) = 0, \\
\lb{4.2b}
& & 2\xi^{2}\left(\mu'' + \nu''\right)
+ \xi\left(\mu' + \nu'\right)
\left[\xi\left(\nu' -2 \mu' - \lambda'\right)
+ 4 \right]\nb\\
& & + e^{\lambda -\nu}\left[2\left(\mu'' - \lambda''\right)
- \nu'\left(\mu' - \lambda'\right)
- \lambda'\left(\mu' + \lambda'\right)\right] = 0,\\
\lb{4.2c}
& &  4\left(\xi^{2}\mu'\right)' 
- \xi^{2}\mu'\left[(3+k)\mu' + 2\left(\lambda' 
-k\nu'\right)\right]\nb\\
& & - e^{\lambda - \nu}\left\{4k\mu'' 
-  \mu'\left[(1+3k)\mu' - 2\left(\lambda' 
-k\nu'\right)\right]\right\} = 0,\\
\lb{4.2d}
 \Phi &=& e^{-\nu}\left[4\mu'' 
-  \mu'\left(3\mu' + 2\nu'\right)\right]
+ e^{-\lambda}\xi^{2}\mu'\left(\mu' -2\nu'\right),\\
\lb{4.2e}
 \Pi &=&  e^{-\lambda}\left[4\left(\xi^{2}\mu'\right)'
- \xi^{2} \mu'\left(3\mu' + 2\lambda'\right)\right]
+ e^{-\nu}\mu'\left(\mu' - 2\lambda'\right),
\eqn
where a prime denotes ordinary differentiation with respect to $\xi$.
In writing Eqs.(\ref{4.2a})-(\ref{4.2e}),   we have omitted
all zero subscripts for simplicity, however,  
the functions $\mu,\; \nu$ and 
$\lambda$ appearing in Eqs.(\ref{4.2a})-(\ref{4.2e}) are functions 
of $\xi$ only. 

Integrating Eq.(\ref{4.2a}), we obtain
\bq
\lb{4.3}
\mu' = \frac{c_{0}}{\xi}e^{(\mu + \nu + \lambda)/2},
\eq
where $c_{0}$ is a non-zero and otherwise arbitrary constant.
Inserting it into Eq.(\ref{4.2c}),   we find that
\bqn
\lb{4.4}
e^{\mu/2} &=& \frac{2e^{-(\nu + \lambda)/2}}
{c_{0}(1+k)\left(\xi^{2}e^{\nu} - e^{\lambda}\right)}\nb\\
& & \times \left\{\xi^{2}e^{\nu}\left[2 + (1+k)\xi \nu'\right]
+ e^{\lambda}\left[2k - (1+k)\xi \lambda'\right]\right\},
\eqn
for $k \not= -1$. When $k = -1$, substituting Eq.(\ref{4.3}) into
Eq.(\ref{4.2c}) and considering Eq.(\ref{4.3})  yields  
\bqn
\lb{4.4a}
\lambda &=& - \frac{\mu}{2} 
          + \ln\left(\frac{\xi^{2}\mu'}{c_{0}}\right),\nb\\
\nu &=& - \frac{\mu}{2} 
+ \ln\left(\frac{\mu'}{c_{0}}\right),\;\;\; (k = -1),
\eqn
where $\mu$ is an arbitrary function of $\xi$ only.
It can be shown that for such solutions Eq.(\ref{4.2b}) is satisfied 
automatically, while Eqs.(\ref{4.2d}) and (\ref{4.2e}) yield
\bq
\lb{4.4b}
\Phi = 0 = \Pi, \;\;\; (k = -1),
\eq
that is, the corresponding spacetime is vacuum. In fact, it is not only 
vacuum but also flat, as the corresponding Riemann tensor  vanishes identically.
Thus, {\em  no self-similar toroidal
solutions of the first kind exist with the equation of state $ p = - \rho$}.
In the following we    consider only the case where $k \not= -1$.

The combination of Eqs.(\ref{4.2a}), (\ref{4.2b}),
(\ref{4.3}) and (\ref{4.4}) yields,
\bqn
\lb{4.5a}
& & \xi\left[\xi^{2}\left(2e^{\lambda} - \xi^{2}e^{\nu}\right)e^{\nu}
- e^{2\lambda}\right] \lambda'\nb\\
& & \;\;\;\;\;\; - k \xi^{4}e^{2\nu}\left[3(1+k)\xi\nu'
+ \xi \lambda' + 4\right]\nb\\
& & \;\;\;\;\;\; + ke^{2\lambda}\left[(1+k)\xi\nu' 
- (5 + 4k)\xi\lambda' + 4(2k + 1)\right]\nb\\
& & \;\;\;\;\;\; +2 k \xi^{2}e^{\nu+\lambda}\left[(1+k)\xi\nu'
+ (2k + 3)\xi\lambda' - 4k\right] = 0,\\
\lb{4.5b}
& & \xi^{4}e^{2\nu}\left\{(1+k)^{2}\xi^{2}\left(2\nu'' - 3{\nu'}^{2}
 - \nu' \lambda'\right) + (1+k)\xi\left[(5k - 8)\nu' - \lambda'\right]
 \right.\nb\\
 & & \;\;\;\;\;\; \left. + 4(k-2)\right\}\nb\\
 & & \;\;\;\;\;\; - \xi^{2}e^{\nu + \lambda}\left\{(1+k)^{2}\xi^{2}\left(2\nu'' 
 + 2\lambda'' + {\nu'}^{2} + {\lambda'}^{2} - 10 {\nu'}{\lambda'}\right)
 \right.\nb\\
 & & \;\;\;\;\;\;\left.
 + 2(1+k)\xi\left[(11k +4)\nu' - 7\lambda'\right]
 + 8(4k+1)\right\}\nb\\
& &  \;\;\;\;\;\;  + e^{2\lambda}\left\{(1+k)^{2}\xi^{2}\left(2\lambda'' 
- 3{\lambda'}^{2} - \nu' \lambda'\right) 
 + (1+k)\xi\left[k\nu' +\left(16k + 3\right)\lambda'\right]\right.\nb\\
 & & \;\;\;\;\;\;\;\;\; \left.
 - 4k(4k+1)\right\} = 0, \;\; (k \not= -1).
\eqn
From the above, we see that the problem of solving the Einstein field
equations now reduces to solving Eqs.(\ref{4.5a}) and (\ref{4.5b}) for the
functions of $\nu$ and $\lambda$. Once they are found, Eqs.(\ref{4.3}),
(\ref{4.2d}) and (\ref{4.2e}) will give the functions $\mu,\; \Phi$ and
$\Pi$.
Eqs.(\ref{4.5a}) and (\ref{4.5b}) are non-linear, and finding 
the solutions of such equations is, in general, very complicated. To start with, let
us  consider the case   $k = 0$.

 \subsection{$k = 0$}

When $k = 0$, Eq.(\ref{4.5a}) reduces to  
\bq
\lb{4.6}
\lambda' \left(\xi^{2}e^{\nu} - e^{\lambda}\right) = 0,
\eq
which has two solutions,
\bq
\lb{4.7}
{\mbox{(i)}} \; \lambda = \lambda_{0},\;\;\;\;
{\mbox{(ii)}} \; \lambda = \nu + 2\ln(\xi),
\eq
where $\lambda_{0}$ is a constant. However, it can be shown that in the latter 
case the corresponding  spacetime is vacuum, $\Phi = 0 = \Pi$. Thus, 
in the following,  we    consider only the first case, for which we can set 
$\lambda_{0} = 0$ without loss of generality. Then, Eq.(\ref{4.5b}) reduces to
\bq
\lb{4.8}
2\xi^{2}\left(\xi^{2}e^{\nu} - 1\right)\nu'' 
- \xi^{2}\left(3\xi^{2}e^{\nu} + 1\right){\nu'}^{2}
- 8\left(\xi^{2}e^{\nu} + 1\right)\left(\xi \nu' + 1\right) = 0.
\eq
To solve this equation, we introduce the function $y(\xi)$ via the
relation
\bq
\lb{4.9}
y \equiv \xi^{2}e^{\nu} - 1,
\eq
for which Eq.(\ref{4.8}) takes the form,
\bq
\lb{4.10}
\frac{y''}{y'} - \frac{5y + 4}{2y(1 + y)}y' = - \frac{2}{\xi}.
\eq
It can be shown that   the general solution is
\bq
\lb{4.11}
\frac{2Y}{Y^{2} - 1} + \ln\left|\frac{Y -1}{Y+1}\right| = \frac{a}{\xi} + b,
\eq
where $a$ and $b$ are two integration constants, and
\bq
\lb{4.12}
Y \equiv \xi e^{\nu/2}.
\eq
In terms of $Y$, Eqs.(\ref{4.2d}) and (\ref{4.4}) yield
\bqn
\lb{4.13}
\Phi &=& {2a}\;\frac{\left(Y^{2} - 1\right)^{2}}{{\xi}Y},\nb\\
e^{\mu} &=& \left(\frac{a}{2c_{0}}\right)^{2}
\left(Y^{2} - 1\right)^{2},\;\; (k = 0).
\eqn

\subsection{$k \not= 0, \; -1$} 

When $\lambda = \nu + 2\ln(\xi)$, spacetime is vacuum for any value of $k$. 
Then, for $k \not= 0, \; -1$ we consider the following ansatz,
\bq
\lb{4.14}
\lambda(\xi) = \nu(\xi) + 2\ln(\xi) + q(\xi),
\eq
where $q(\xi)$ is an arbitrary function of $\xi$. Substituting 
into Eq.(\ref{4.5a}) we obtain
\bq
\lb{4.15}
\left(1 + 3k\right) \nu' 
+ \left( 1 - \frac{4k}{1 - e^{q}}e^{q}\right)q' 
+ \frac{2(1+3k)}{(1+k)\xi} = 0.
\eq
To solve this equation, let us first consider the case $k = -1/3$.

\subsubsection{$k = -1/3$}

In this case Eq.(\ref{4.15}) has the solution,
\bq
\lb{4.16}
q(\xi) = q_{0}, 
\eq
where $q_{0}$ is a constant. Inserting Eqs.(\ref{4.14}) and (\ref{4.16})
into Eq.(\ref{4.5b}),  we find 
\bq
\lb{4.17}
\xi^{2}\nu'' - 2\xi^{2}{\nu'}^{2} - 9\xi\nu'
- 12 = 0.
\eq
There are two particular solutions,
\bq
\lb{4.18}
(1) \;\; \nu_{s}^{(1)} = - 2\ln(\xi), \;\;\;\;\;
(2) \;\; \nu_{s}^{(2)} = - 3\ln(\xi).
\eq
To find the general solution of Eq.(\ref{4.17}) we set
$\nu = A(\xi) + \nu_{s}^{(1)}$, and then Eq.(\ref{4.17})
reduces to
\bq
\lb{4.19}
A'' - 2{A'}^{2} - \frac{1}{\xi}A' = 0,
\eq
which has the general solution,
\bq
\lb{4.20}
A(\xi) = - \frac{1}{2}\ln\left|A_{1} - A_{0}\xi^{2}\right|,
\eq
where $A_{0}$ and $A_{1}$ are two integration constants. Thus, 
for $k = -1/3$  we have the following general solutions,
\bqn
\lb{4.21}
\lambda(\xi) &=& q_{0} 
- \frac{1}{2}\ln\left|A_{1} - A_{0}\xi^{2}\right|,\nb\\
\nu(\xi) &=&   
- \frac{1}{2}\ln\left|\xi^{4}\left(A_{1} - A_{0}\xi^{2}\right)\right|,\nb\\
e^{\mu/2} &=&  \frac{2A_{1}e^{-q_{0}/2} \xi}
              {c_{0}\left(A_{1} - A_{0}\xi^{2}\right)^{1/2}},\nb\\
\Phi &=& -3\Pi = \frac{12A_{1}\left(1 - e^{q_{0}}\right)}
              {e^{q_{0}}\left(A_{1} - A_{0}\xi^{2}\right)^{1/2}}.
\eqn

\subsubsection{$k \not= -1/3$}

In this case, Eq.(\ref{4.15}) has the solution,
\bq
\lb{4.22}
\nu(\xi) = - \frac{2}{1+k}\ln(\xi) - \frac{1}{1+3k}
\left[q + 4k\ln\left(1 - e^{q}\right)\right]
+ \nu_{0}, 
\eq
where $\nu_{0}$ is a constant. Inserting the above expression
into Eq.(\ref{4.5b}),   we find 
\bq
\lb{4.23}
2 f(q)q'' - g(q){q'}^{2} - \frac{4}{\xi} h(q) q' = 0,
\eq
where
\bqn
\lb{4.24}
f(q) &\equiv& \left(3k^{4} + 10 k^{3} + 12k^{2} + 6k +1\right)
\left(1 - e^{3q}\right)\nb\\
& & + \left(6k^{5} + 23 k^{4} + 34 k^{3} + 24k^{2} + 8k +1\right)e^{2q}\nb\\
& & - \left(3k^{5} + 16 k^{4} + 32 k^{3} + 30k^{2} + 13k +2\right)e^{q},\nb\\
g(q) &\equiv& k\left(7k^{4} + 22 k^{3} + 24k^{2} + 10k +1\right)e^{3q} \nb\\
& & - \left(10k^{5} + 37 k^{4} + 50 k^{3} + 28k^{2} + 4k -1\right)e^{2q}\nb\\
& & + \left(3k^{5} + 12 k^{4} + 20 k^{3} + 18k^{2} + 9k +2\right)e^{q}\nb\\
& & - 3(1+k)^{2}\left(1- k^{2}\right),\nb\\
h(q) &\equiv& k\left(3k^{3} + 7k^{2} + 5k +1\right)e^{3q} \nb\\
& & - \left(6 k^{4} + 17 k^{3} + 17k^{2} + 7k +1\right)e^{2q}\nb\\
& & + \left(3k^{4} + 13 k^{3} + 19 k^{2} + 11k +2\right)e^{q}\nb\\
& & - \left(3 k^{3} + 7 k^{2} + 5k +1\right).
\eqn
It can be shown that when $q'(\xi) = 0$ the corresponding solutions represent
a vacuum space.  When $q' \not= 0$, Eq.(\ref{4.23}) can be written as
\bq
\lb{4.26}
2 F(q)\left[\ln\left(q'\right)\right]'  - G(q)  q' = \frac{4}{\xi},
\eq 
where
\bq
\lb{4.27}
F(q) \equiv \frac{f(q)}{h(q)},\;\;\;
G(q) \equiv \frac{g(q)}{h(q)}.
\eq
Solving Eq.(\ref{4.26}) is not a trivial exercise. One may first 
try to find a particular solution of it, say,  $q_{s}(\xi)$. Once such a 
solution is known, setting $q = q_{s}(\xi) + q_{0}(\xi)$, we find that
Eq.(\ref{4.26})  reduces to the following form  for $q_{0}(\xi)$,
\bq
\lb{4.28}
\left[\ln\left({q_{0}}'\right)\right]'  - H(q_{0},q_{s})  {q_{0}}' = 0.
\eq 
Unfortunately, we have not yet been able to find such a particular solution.


\section{Physical and Geometrical Interpretations of the Self-Similar
Solutions}
\renewcommand{\theequation}{5.\arabic{equation}}
 \setcounter{equation}{0}
 
In this section, we    study the local, as well as the global, properties of the 
self-similar solutions for $k = 0$ and $-1/3$, obtained in the last section. Note that,
although these solutions were found by taking the limit
$\eta \rightarrow 0$, in this section we    extend them to any   $t \in
(-\infty, 0)$.  

\subsection{Self-Similar Solutions With $k = 0$}

These are the solutions given  by Eqs.(\ref{4.7})-(\ref{4.13}).   Rescaling the 
coordinates $t$ and $z$ and using the conformal transformation ${g'}_{\mu\nu}
= B^{2}g_{\mu\nu}$, where $B$ is a constant,   without loss of
generality, we  set 
\bq
\lb{5.1}
c_{0} = \frac{1}{2}a, \;\;\; A_{1} = 1,
\eq
for which the metric reads
\bq
\lb{5.2}
ds^{2} = dt^{2} - \frac{Y^{2}}{\xi^{2}}dz^{2} 
- \frac{{L_{0}}^{2}}{\left(Y^{2}-1\right)^{2}}
\left(d\theta^{2} + d\varphi^{2}\right).
\eq
Then, the corresponding energy density is given by
\bq
\lb{5.3}
\rho = \frac{1}{4t^{2}}\Phi = \frac{Y^{2}-1}{2t^{2}Y} I(Y),
\eq
where
\bqn
\lb{5.4a}
I(Y) &\equiv& - b(Y^{2} - 1) + 2Y + (Y^{2} -1)\ln\left|\frac{Y-1}{Y+1}\right|,\\
\lb{5.4b}
\xi(Y) &=& \frac{a(Y^{2}-1)}{I(Y)}.
\eqn
From the above expressions one can see that the spacetime is singular at
$t = 0$ and $\xi = -a/b$, where
\bq
\lb{5.4c}
\xi(Y) = - \frac{a}{b}, \;\;\; {\mbox{when}} \; Y = 0, \; \pm \infty.
\eq
The singularity at $t= 0$ is always spacelike, while the nature of the singularity
at $\xi = -a/b$ depends on the values of $Y$. The normal vector
to the surface $\xi = -a/b$ is given by
\bq
\lb{5.4d}
n_{\lambda} \equiv \frac{\partial}{\partial x^{\lambda}}\left(z - \frac{a}{b}t\right)
= - \frac{a}{b}\delta^{t}_{\lambda} + \delta^{z}_{\lambda}.
\eq
Then,  
\bq
\lb{5.4e}
n_{\lambda} n^{\lambda} = \frac{a^{2}}{b^{2}} - \frac{\xi^{2}}{Y^{2}}
= \cases{a^{2} > 0, & $Y = \pm \infty$,\cr
-\infty < 0, & $Y = 0$.\cr}
\eq
Therefore, when $Y = 0$ the corresponding spacetime singularity at $x = -a/b$ is timelike,
and when $Y = \pm \infty$, it is spacelike. 
From Eqs.(\ref{B.3}) and (\ref{5.2}) we obtain
\bqn
\lb{5.4f}
{\cal R} &=& L_{0}e^{-\mu/2} = \pm \frac{L_{0}}{Y^{2} - 1},\nb\\
{\cal R}_{,\lambda} &=& \pm \frac{2L_{0}YY_{,\xi}}{t\left(Y^{2} - 1\right)^{2}}
\left(\xi\delta^{t}_{\lambda} + \delta^{t}_{\lambda}\right).
\eqn
Thus, we have 
\bq
\lb{5.4g} 
{\cal R}_{,\lambda} {\cal R}^{,\lambda} = \frac{L_{0}^{2}I^{2}(Y)}{4t^{2}
\left(Y^{2} - 1\right)}.
\eq
From Eq.(\ref{5.4b}) we also find that
\bq
\lb{5.5}
\frac{d\xi(Y)}{dY} = \frac{4a}{I^{2}(Y)},
\eq
which shows that $\xi(Y)$ is a monotonically increasing ($a>0$) or decreasing 
($a<0$) function of $Y$, depending on the sign of the constant $a$. To have 
the energy density $\rho$ non-negative, the solutions are restricted to the 
following regions for different $a$,
\bq
\lb{5.6}
Y \xi  = \cases{\ge 0, & $a > 0$,\cr
\le 0, & $a < 0$,\cr}
\eq
as one can see clearly from Eq.(\ref{4.13}). To further study  the properties of the
solutions, it is convenient to  consider the following four cases separately:
(a) $\; a > 0, \; b > 0$; (b) $\; a > 0, \; b < 0$; (c) $\; a < 0, \; b > 0$; and (d) 
$\; a < 0, \; b < 0$.

{\bf Case (a) $\; a > 0, \; b > 0$:} In this case we find that  
\bq
\lb{5.7}
I(Y) = \cases{ -\infty, & $Y = \pm \infty$, \cr
 b, & $Y = 0$, \cr
 \pm 2, & $Y = \pm 1$, \cr}
\eq
and $I(Y) = 0$ has two real roots, $Y_{\pm}$, with the properties
\bq
\lb{5.8}
Y_{+} > 1, \;\;\; -1 < Y_{-} < 0,
\eq
as  can be seen from Fig. 1. 

 \begin{figure}[htbp]
 \begin{center}
 \label{fig1}
 \leavevmode
  \epsfig{file=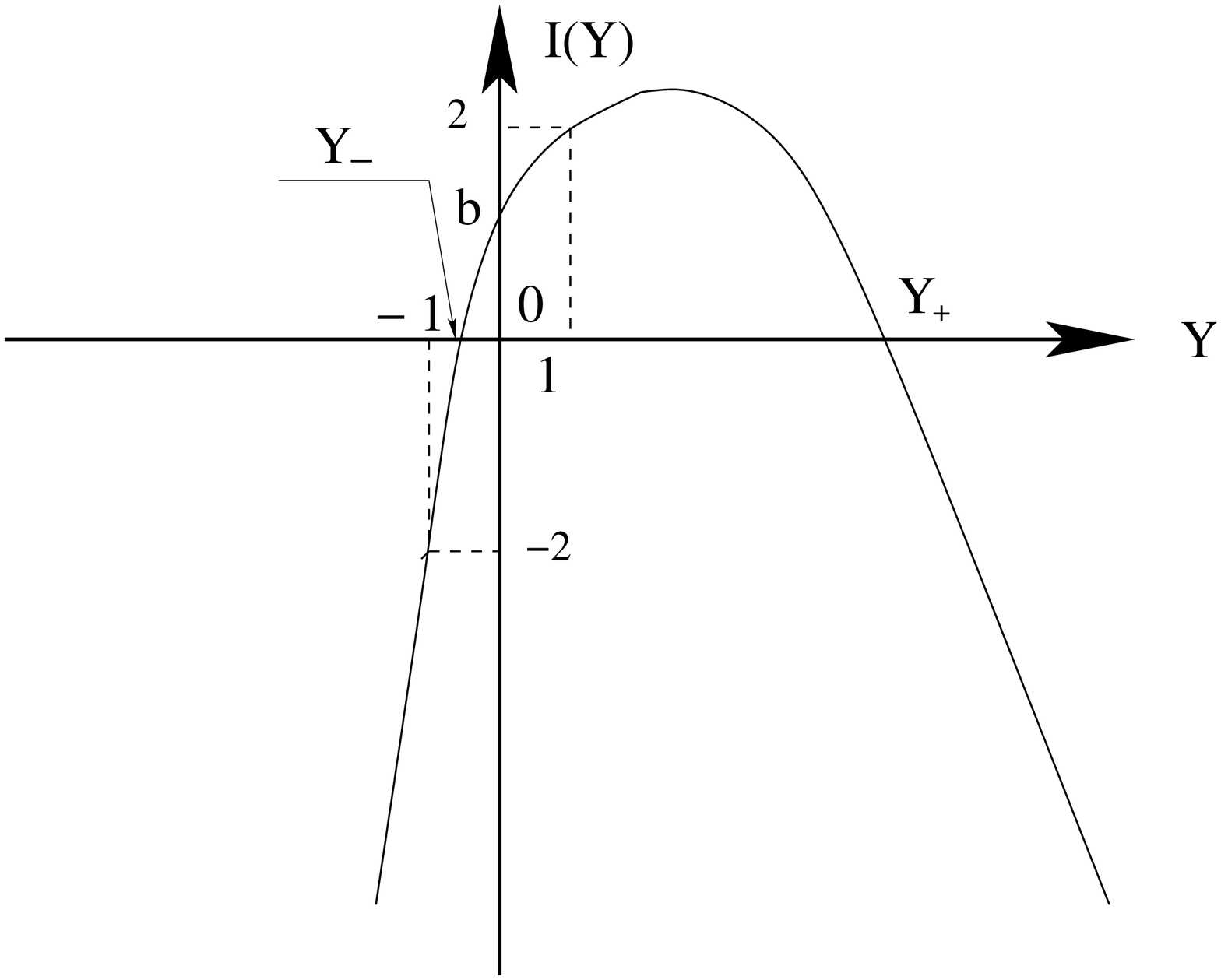,width=0.45\textwidth,angle=0}
 \caption{The function $I(Y)$ versus $Y$ for $a > 0$ and $b> 0$.}
 \end{center}
 \end{figure}

From Eqs.(\ref{5.4a}) and (\ref{5.4b}) we find that
 \bq
 \lb{5.9}
\xi(Y) = \cases{ -a/b, & $Y = 0, \; \pm \infty$, \cr
 0, & $Y = \pm 1$,\cr
 \infty, & $Y = Y_{\pm}$. \cr} 
\eq
Then, one can see that $\xi(Y)$ must behave as that given by Fig. 2, from which, 
together with Eq.(\ref{5.6}), we can see that the energy density is non-negative only
in the following three regions,
\bqn
\lb{5.10}
& & (i) \;\; Y \in [-\infty,\; -1], \;\;\; {\mbox{or}} \;\;\;
\xi \in [-{a}/{b}, \; 0], \nb\\
& & (ii) \;\; Y \in (Y_{-},\; 0], \;\;\; {\mbox{or}} \;\;\;
\xi \in (-\infty, \; 0], \nb\\
& & (iii) \;\; Y \in [1,\; Y_{+}), \;\;\; {\mbox{or}} \;\;\;
\xi \in [0, \; \infty).
\eqn

 \begin{figure}[htbp]
 \begin{center}
 \label{fig2}
 \leavevmode
  \epsfig{file=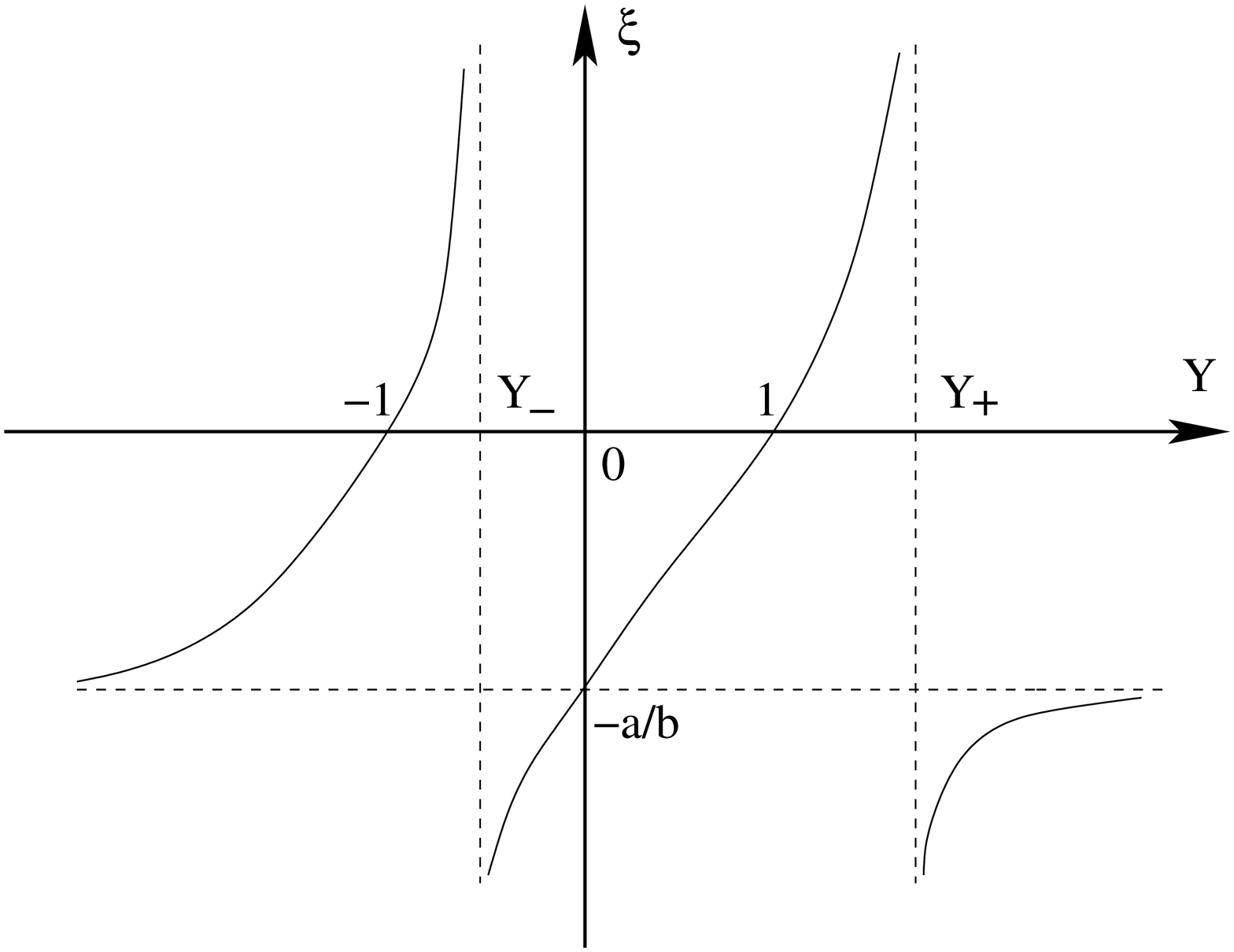,width=0.45\textwidth,angle=0}
 \caption{The function $\xi(Y)$ versus $Y$ for $a > 0$ and $b> 0$.}
 \end{center}
 \end{figure}
 
In the region $Y \in [-\infty,\; -1]$ or  $\xi \in [-{a}/{b}, \; 0]$, the
spacetime is singular at $\xi = -a/b$ or $Y = -\infty$. From Eq.(\ref{5.4e})
we can see that this singularity is spacelike. In the ($t,z$)-plane, this region
is   between the two lines $\xi = -a/b$ and $\xi = 0 (z = 0)$, 
as shown in Fig. 3, in which it is referred to as Region $I$.
The metric is singular at $\xi = 0$ or $Y = -1$. However, this 
singularity is a coordinate one, as one can see from the expression of
$\rho$, which is finite there. The nature of this
surface is null. In fact, introducing the normal vector $N_{\lambda}$ to 
this surface by
\bq
\lb{5.11}
N_{\lambda} \equiv \frac{\partial z}{\partial x^{\lambda}} = \delta^{z}_{\lambda},
\eq
we find that
\bq
\lb{5.12}
N_{\lambda}  N^{\lambda} = \frac{\xi^{2}}{Y^{2}} \rightarrow 0,
\eq
as $\xi \rightarrow 0$ and $Y \rightarrow -1$. On the other hand, from
Eq.(\ref{5.4g}) we find that in this region ${\cal R}_{,\lambda}$ is always
timelike, that is, the whole region is trapped.

Region $II$, where $Y \in (Y_{-},\; 0]$ or $
\xi \in (-\infty, \; 0]$, is the region in between the two spacetime singularities
$t = 0$ and $\xi = -a/b$. In this region,  ${\cal R}_{,\lambda}$ is always
spacelike, that is, this region is untrapped. However, because of the two spacetime
singularities, it is difficult to interpret this region physically.

In Region $III$, where  $Y \in [1,\; Y_{+})$, or $\xi \in [0, \; \infty)$, the
vector ${\cal R}_{,\lambda}$ is always timelike, that is, all of this region is 
trapped. The spacetime is singular at $t = 0$.

 \begin{figure}[htbp]
 \begin{center}
 \label{fig3}
 \leavevmode
  \epsfig{file=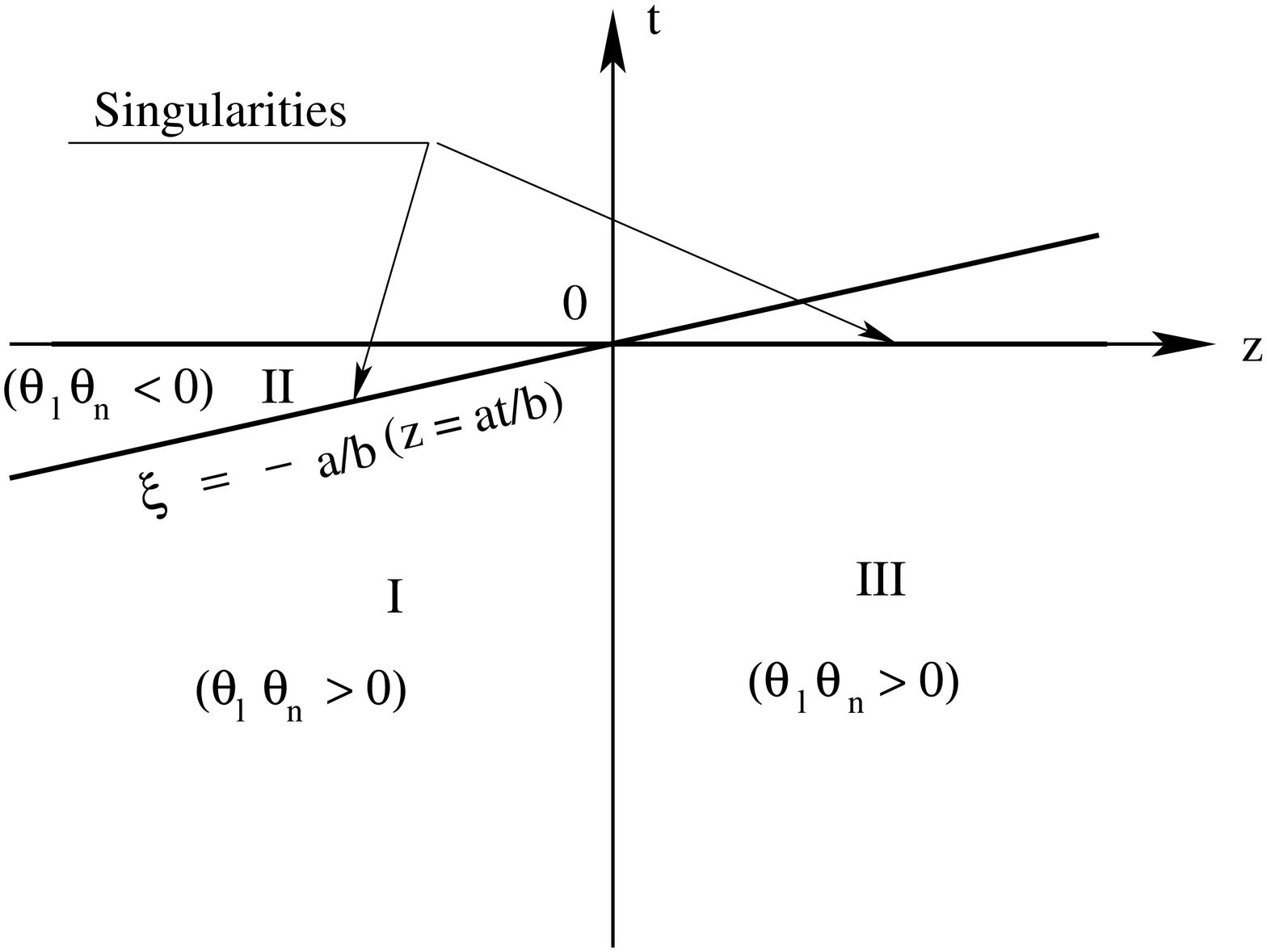,width=0.45\textwidth,angle=0}
 \caption{The ($t, \; z$)-plane for $a > 0$ and $b> 0$. The spacetime is singular 
 on the lines $\xi = -a/b$ and $t = 0$. }
 \end{center}
 \end{figure}

{\bf Case (b) $\; a > 0, \; b < 0$:} In this case we find
\bq
\lb{5.13}
I(Y) = \cases{ \infty, & $Y =   \pm \infty$, \cr
 -|b|, & $Y = 0  $,\cr
 \pm 2, & $Y =  {\pm} 1$, \cr} 
\eq
for which $I(Y) = 0$ has also two real roots, $Y_{\pm}$, but now with
\bq
\lb{5.14}
0< Y_{+} < 1, \;\;\; Y_{-} < -1,
\eq
as shown in Fig. 4.

 \begin{figure}[htbp]
 \begin{center}
 \label{fig4}
 \leavevmode
  \epsfig{file=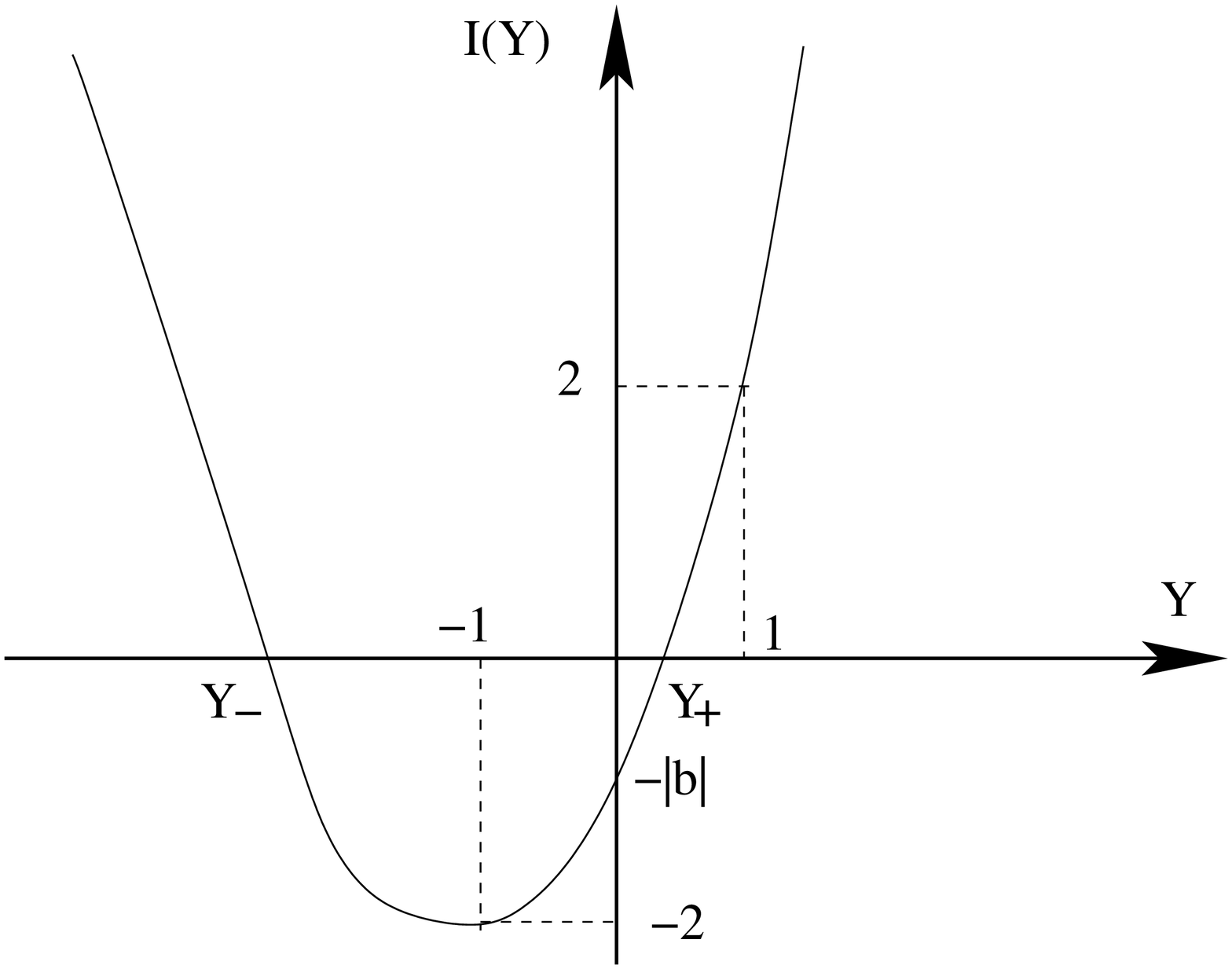,width=0.45\textwidth,angle=0}
 \caption{The function $I(Y)$ versus $Y$ for $a > 0$ and $b < 0$.}
 \end{center}
 \end{figure}

Then, it can be shown that  
 \bq
 \lb{5.15}
\xi(Y) = \cases{ a/|b|, & $Y = 0, \; \pm \infty$, \cr
 0, & $Y = \pm 1$,\cr
 \infty, & $Y = Y_{\pm}$, \cr} 
\eq
and the curve of $\xi(Y)$ versus $Y$ is given by Fig. 5, from which we can see that
the energy density is non-negative only
in the  regions,
\bqn
\lb{5.16}
& & (i) \;\; Y \in (Y_{-},\; -1], \;\;\; {\mbox{or}} \;\;\;
\xi \in (-\infty, \; 0], \nb\\
& & (ii) \;\; Y \in [1,\; \infty], \;\;\; {\mbox{or}} \;\;\;
\xi \in [0, \; a/|b|].
\eqn

 \begin{figure}[htbp]
 \begin{center}
 \label{fig5}
 \leavevmode
  \epsfig{file=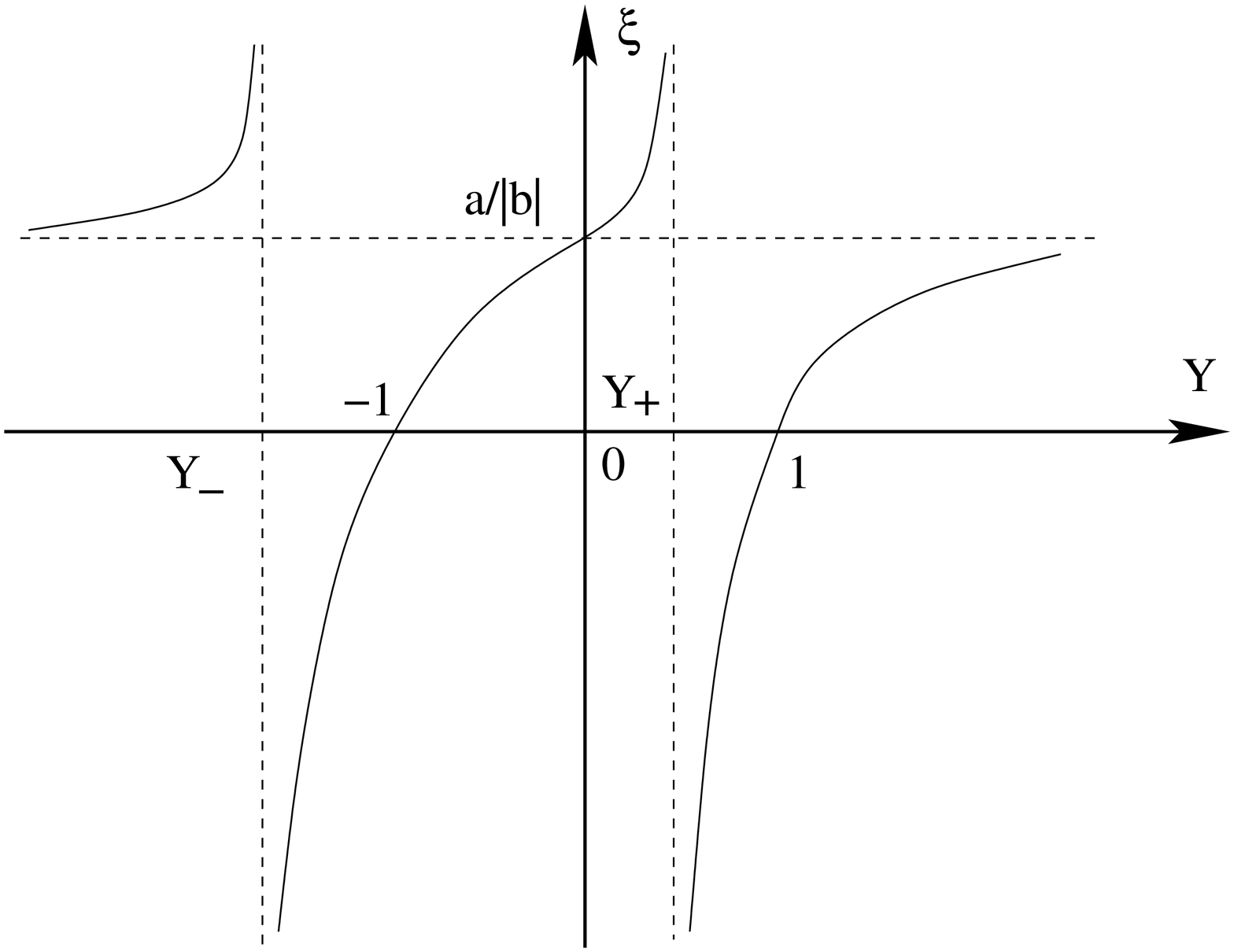,width=0.45\textwidth,angle=0}
 \caption{The function $\xi(Y)$ versus $Y$ for $a > 0$ and $b< 0$.}
 \end{center}
 \end{figure}
 
In the region $Y \in (Y_{-},\; -1]$ or  $\xi \in (-\infty, \; 0]$, the
spacetime is singular at $t = 0$ or $Y = Y_{-}$.  In the ($t,z$)-plane, this region
is the one where $t \le 0$ and $z \le 0$, marked as Region $I$ in Fig. 6. 
The metric is singular at $\xi = 0$ or $z = 0$. As in the previous case,
this singularity is a coordinate one,  and the  surface is null, as
shown by Eq.(\ref{5.11}).   Moreover, in this region ${\cal R}_{,\lambda}$ is always
timelike, except for the line $t = 0$ or $Y = Y_{-}$. Thus, in the present
case the whole region is trapped.

In Region $II$, where $Y \in [1, \; \infty]$ or $
\xi \in [0, \; a/|b|]$,  the spacetime is singular on the line $\xi = a/|b|$
or $z = -at/|b|$, and the nature of the singularity is spacelike, as can be seen
from Eq.(\ref{5.4e}), considering the fact that now $\xi = a/|b|$ corresponds to
$Y= \infty$. In addition, ${\cal R}_{,\lambda}$ is also  
timelike, that is,  Region $II$ in the present case is also trapped.
 
 \begin{figure}[htbp]
 \begin{center}
 \label{fig6}
 \leavevmode
  \epsfig{file=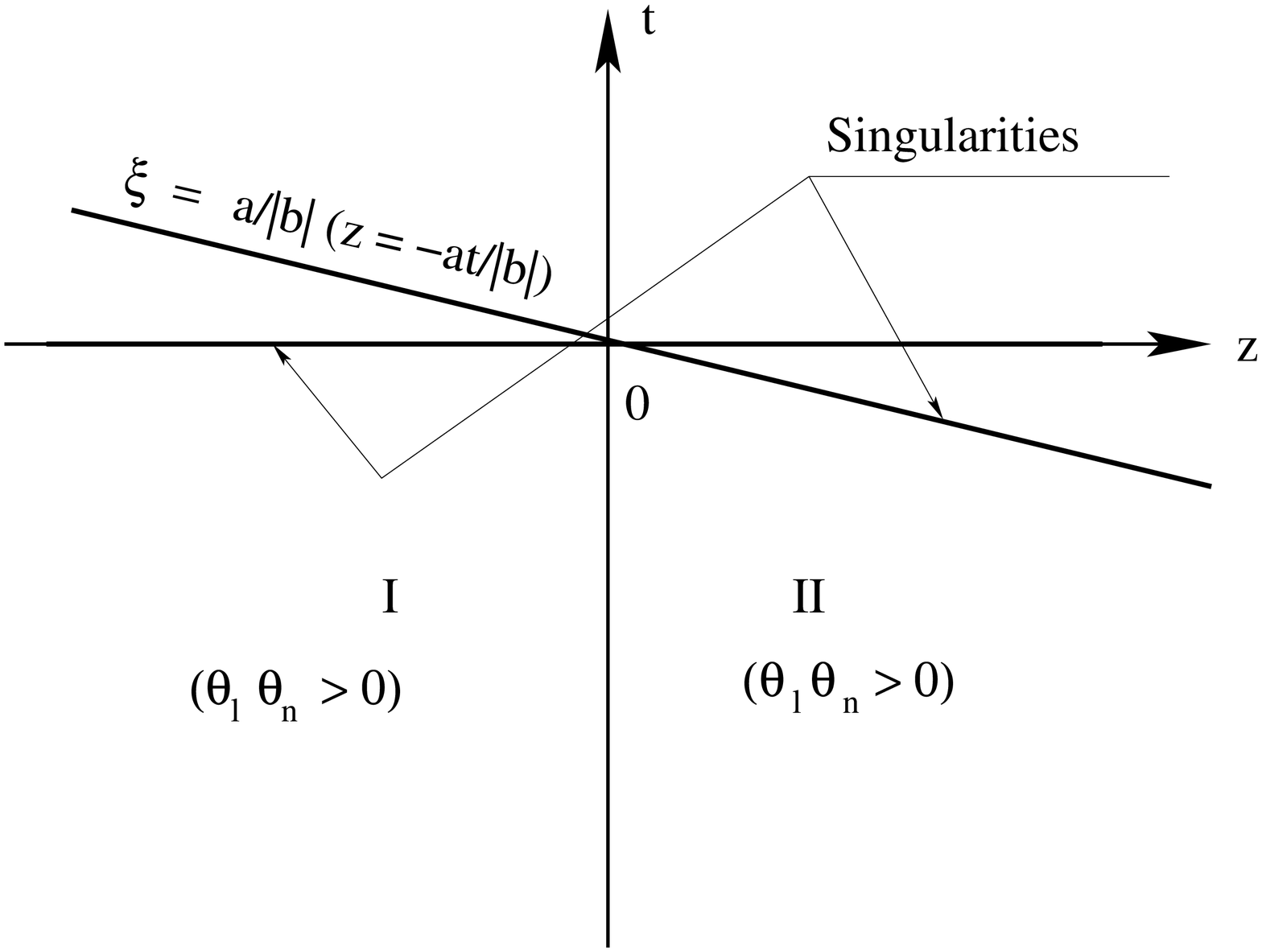,width=0.45\textwidth,angle=0}
 \caption{The ($t, \; z$)-plane for $a > 0$ and $b < 0$. The spacetime is singular 
 on the lines $\xi = a/|b|$ and $t = 0$. }
 \end{center}
 \end{figure}

{\bf Case (c) $\; a < 0, \; b > 0$:} In this case $I(Y)$ is given by
Fig. 1,  and  $\xi(Y)$  behaves as that given by Fig. 7, from which we can see
that the energy density is non-negative only
in the  regions,
\bqn
\lb{5.17}
& & (i) \;\; Y \in [-\infty,\; -1], \;\;\; {\mbox{or}} \;\;\;
\xi \in [{|a|}/{b}, \; 0], \nb\\
& & (ii) \;\; Y \in (Y_{-},\; 0], \;\;\; {\mbox{or}} \;\;\;
\xi \in [|a|/b, \; \infty), \nb\\
& & (iii) \;\; Y \in [1,\; Y_{+}), \;\;\; {\mbox{or}} \;\;\;
\xi \in (-\infty,   \; 0].
\eqn
 \begin{figure}[htbp]
 \begin{center}
 \label{fig7}
 \leavevmode
  \epsfig{file=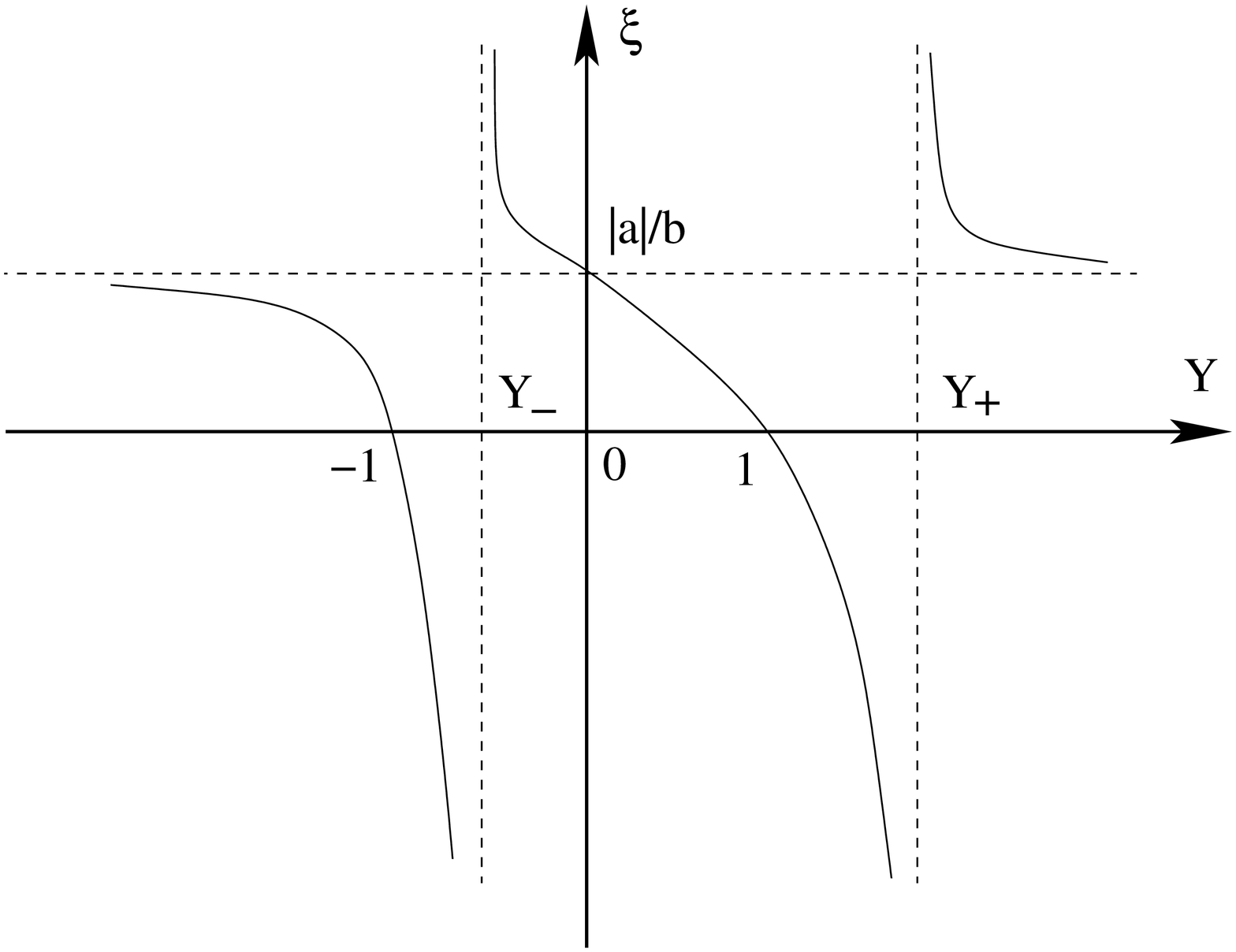,width=0.45\textwidth,angle=0}
 \caption{The function $\xi(Y)$ versus $Y$ for $a < 0$ and $b > 0$.}
 \end{center}
 \end{figure}

The corresponding three regions of Eq.(\ref{5.17}) 
in the ($t, \; z$)-plane are shown in Fig. 8.
In particular, the spacetime is singular on the lines $t = 0$ and $\xi = |a|/b$,
and Regions $I$ and $III$ are trapped, while Region $II$ is not. The metric
is singular on the line $z = 0$ or $Y = \pm 1$, which is a null surface. 

 \begin{figure}[htbp]
 \begin{center}
 \label{fig8}
 \leavevmode
  \epsfig{file=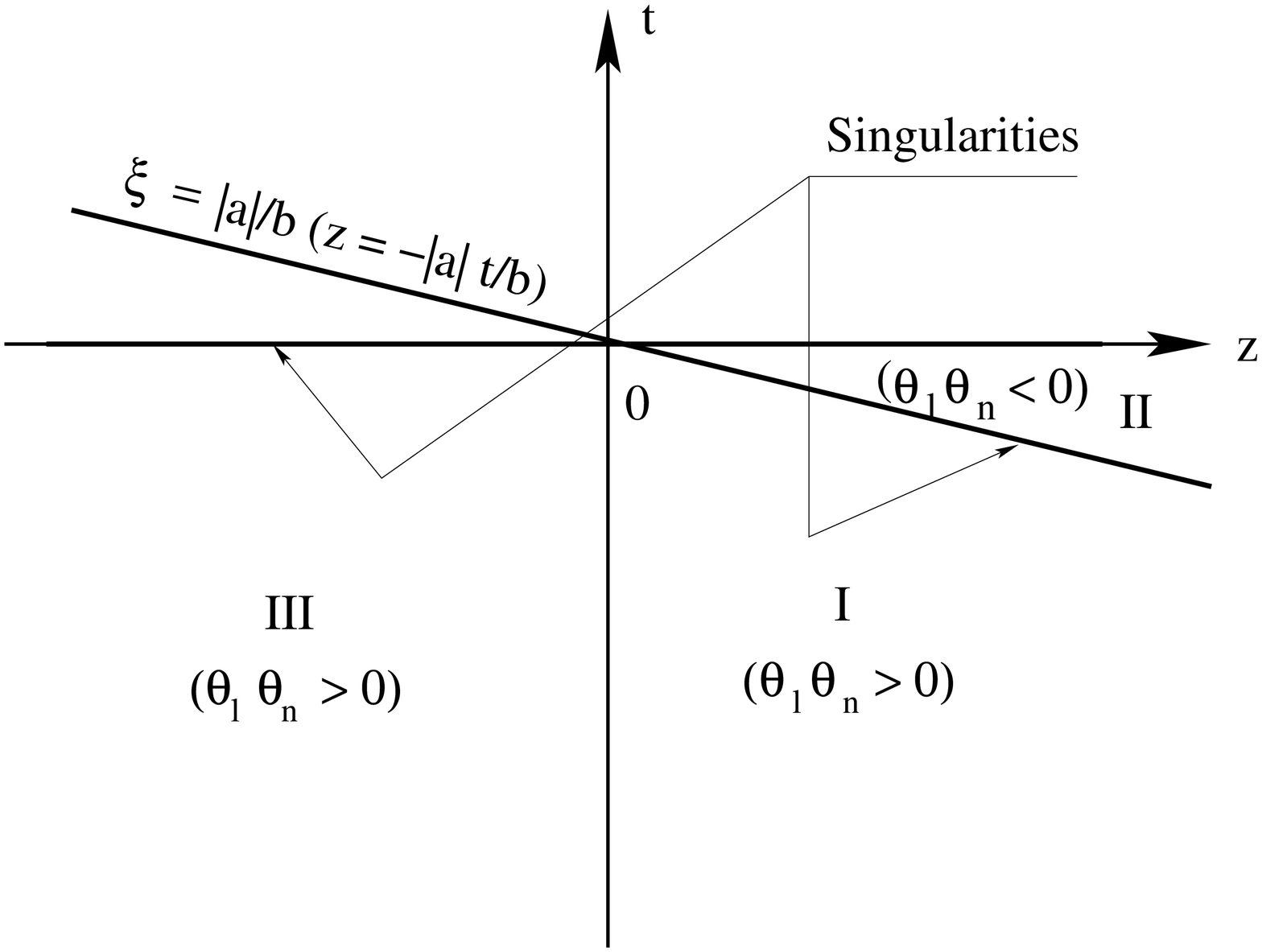,width=0.45\textwidth,angle=0}
 \caption{The ($t, \; z$)-plane for $a < 0$ and $b > 0$. The spacetime is singular 
 on the lines $\xi = |a|/b$ and $t = 0$. }
 \end{center}
 \end{figure}

{\bf Case (d) $\; a < 0, \; b < 0$:} In this case $I(Y)$ is given by
Fig. 4,  and  $\xi(Y)$  behaves as that given in Fig. 9, from which we can see
that the energy density is non-negative only in the  regions,
\bqn
\lb{5.18}
& & (i) \;\; Y \in (Y_{-},\; -1], \;\;\; {\mbox{or}} \;\;\;
\xi \in [0, \; \infty), \nb\\
& & (ii) \;\; Y \in [0, \; Y_{+}), \;\;\; {\mbox{or}} \;\;\;
\xi \in (-\infty, \; - |a/b|],   \nb\\
& & (iii) \;\; Y \in [1,\; \infty], \;\;\; {\mbox{or}} \;\;\;
\xi \in [-|a/b|, \; 0]. 
\eqn
 \begin{figure}[htbp]
 \begin{center}
 \label{fig9}
 \leavevmode
  \epsfig{file=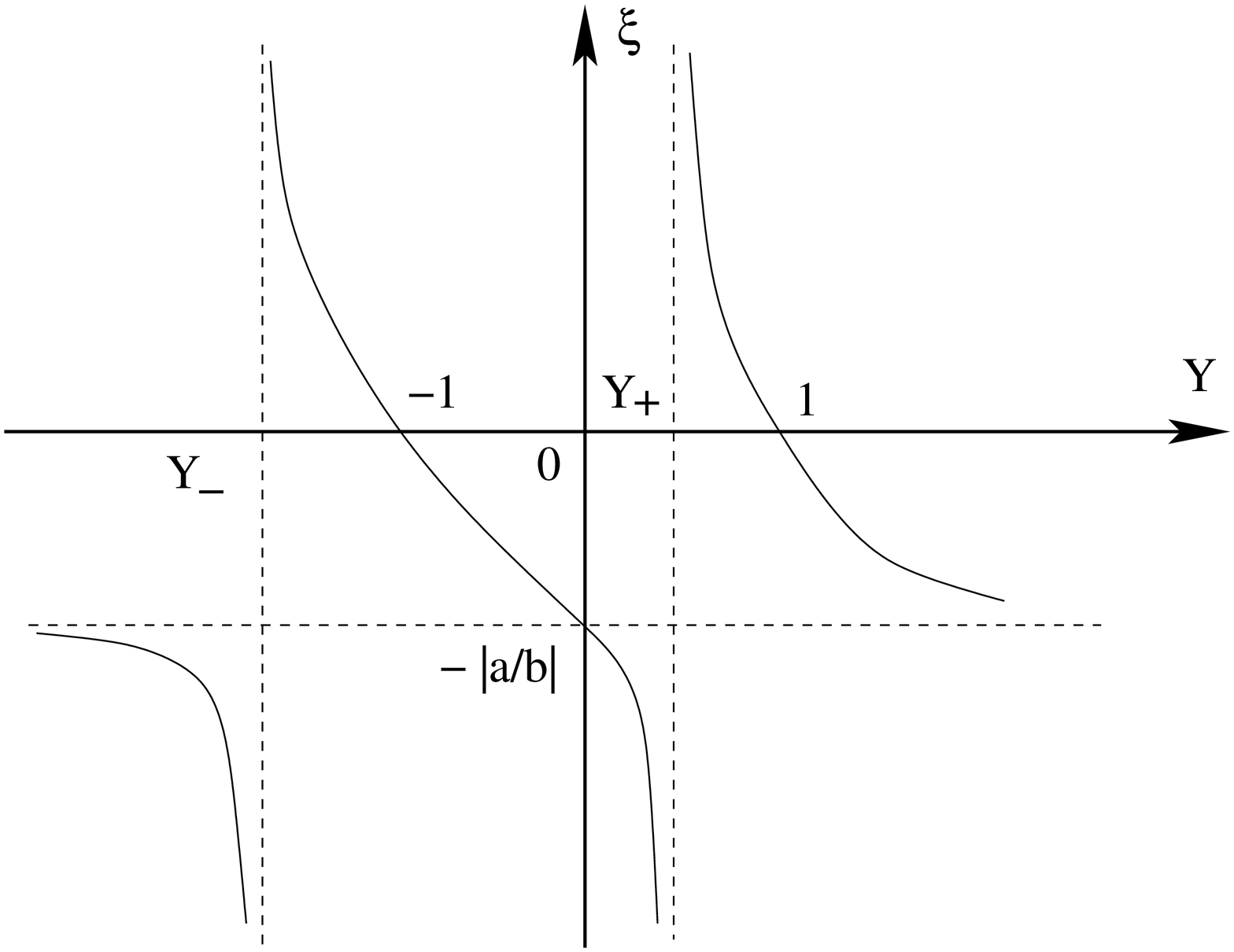,width=0.45\textwidth,angle=0}
 \caption{The function $\xi(Y)$ versus $Y$ for $a < 0$ and $b < 0$.}
 \end{center}
 \end{figure}

The corresponding three regions of Eq.(\ref{5.18}) in the ($t, \; z$)-plane are shown 
in Fig. 10, where the spacetime is also singular on the lines $t = 0$ and $\xi = -|a/b|$,
and Regions $I$ and $III$ are trapped, while Region $II$ is not. The metric
is singular on the line $z = 0$ or $Y = \pm 1$, and this line is null. 

 \begin{figure}[htbp]
 \begin{center}
 \label{fig10}
 \leavevmode
  \epsfig{file=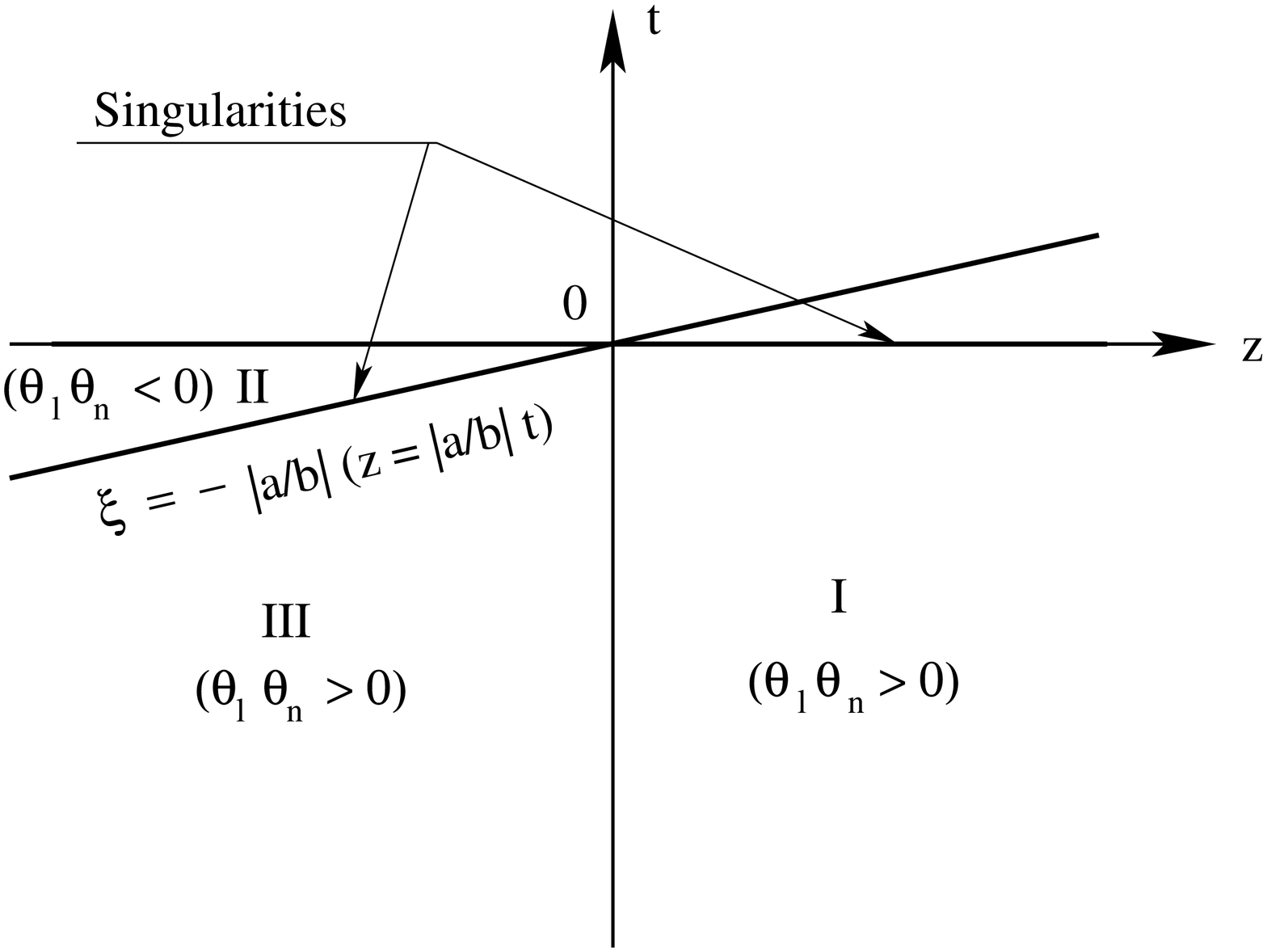,width=0.45\textwidth,angle=0}
 \caption{The ($t, \; z$)-plane for $a < 0$ and $b < 0$. The spacetime is singular 
 on the lines $\xi = |a/b|$ and $t = 0$. }
 \end{center}
 \end{figure}

\subsection{Self-Similar Solutions With $k = -1/3$}

These are the solutions given  by Eq.(\ref{4.21}).   Without loss of
generality, we   set 
\bq
\lb{5.19}
A_{1} = 1, \;\; c_{0} = 2e^{-q_{0}/2}.
\eq
Then, the solutions are given by
\bqn
\lb{5.20}
\lambda &=& q_{0} - \frac{1}{2}\ln\left(1 - \alpha \xi^{2}\right),\nb\\
\nu &=&   - \frac{1}{2}\ln\left[\xi^{4}\left(1 - \alpha \xi^{2}\right)\right],\nb\\
\mu &=&    \ln\left(\frac{\xi^{2}}{1 - \alpha \xi^{2}}\right),
\eqn
for which we have
\bq
\lb{5.21}
\rho = -3p = \frac{\rho_{0}}{t^{2}\left(1 - \alpha \xi^{2}\right)^{1/2}},
\eq
where $\rho_{0} \equiv (1-e^{q_{0}})/(4e^{q_{0}})$ and $\alpha \equiv A_{0}$.
Thus, when $\alpha > 0$, the solutions are valid only in the region where 
$|\xi| \le \xi_{0}$ as shown in Fig. 11, and the spacetime is singular at 
$t = 0$ and at $\xi = \pm \xi_{0} \equiv \pm \alpha^{-1/2}$. These singularities 
are all spacelike. This can be seen easily when $t = 0$. For the cases 
$\xi = \pm \xi_{0}$, we first introduce the normal vectors $n^{\pm}_{\mu}$ by
\bq
\lb{5.22}
n^{\pm}_{\mu} \equiv \frac{\partial(z \pm \xi_{0} t)}{\partial x^{\mu}}
= \delta^{z}_{\mu} \pm \xi_{0}\delta^{t}_{\mu}.
\eq
Then, we find that
\bq
\lb{5.23}
n^{\pm}_{\mu} n^{\pm}_{\nu} g^{\mu\nu} 
= e^{-(\nu + q_{0})}\left(\frac{\xi_{0}^{2}}{\xi^{2}}
- e^{q_{0}}\right) \sim e^{-(\nu(\xi_{0}) + q_{0})}
\left(1- e^{q_{0}}\right) > 0.
\eq
Thus, the singularities on $\xi = \pm \xi_{0}$ are indeed spacelike.
On the other hand, it can be shown that now we have
\bqn
\lb{5.24}
{\cal R} &=& L_{0} \frac{\left(1 - \alpha \xi^{2}\right)^{1/2}}{|\xi|},\nb\\
{\cal R}_{,\alpha}{\cal R}^{,\alpha}
&=& \frac{L_{0}^{2} \xi_{0}e^{-q_{0}}}{t^{2} \xi^{2}
\left(\xi_{0}^{2}/\xi^{2} -1\right)^{1/2}}\left(1 - e^{q_{0}}\right) > 0.
\eqn
Thus, the whole region $ - x_{0} < \xi < x_{0}$ is trapped [cf. Fig. 11].
It should be noted that the metric is singular on $z = 0$ or $\xi = 0$,
but it can be shown that it is a coordinate one. 

 \begin{figure}[htbp]
 \begin{center}
 \label{fig11}
 \leavevmode
  \epsfig{file=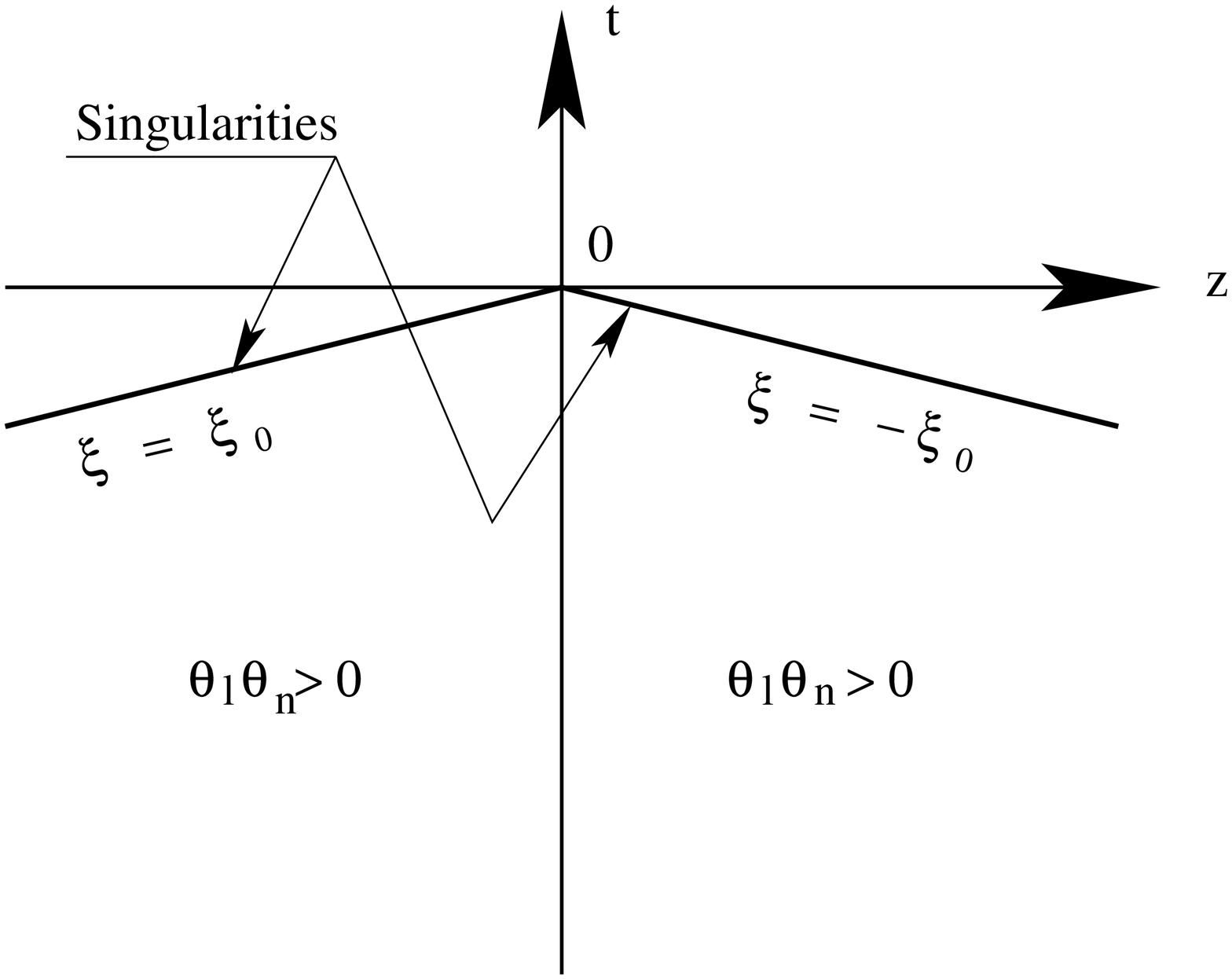,width=0.45\textwidth,angle=0}
 \caption{The spacetime in the ($t, \; z$)-plane for $\alpha > 0$, for which
 the solutions are valid only in the region $|\xi| \le \xi_{0}$. The spacetime 
 is singular on the lines   $\xi = \pm \xi_{0}$, which are
 always spacelike. For $\alpha \le 0$, the 
 solutions are valid in the whole half plane $t \le 0$, and the spacetime 
 is singular only on the line $t = 0$.}
 \end{center}
 \end{figure}

When $\alpha \le 0$, the solutions are valid in the whole half plane
$t \le 0$, and the spacetime is singular only on $t = 0$. It can also be shown
that the whole region $t \le 0$ is now trapped, as one can show that 
$\theta_{l}\theta_{n} > 0$ for all time $t < 0$.

\section{Conclusions}
\renewcommand{\theequation}{6.\arabic{equation}}
 \setcounter{equation}{0}

In this paper, we have studied the asymptotics of solutions of a perfect fluid  
when coupled with a cosmological constant in four-dimensional spacetimes with 
toroidal symmetry. We found that the problem for self-similar solutions of the 
first kind   with  the equation of state, $p = k \rho$, can be 
reduced to solving a master equation of the form,
$$
2 F(q, k)\frac{q''(\xi)}{q'(\xi)}   - G(q,k)  q'(\xi) = \frac{4}{\xi}.
$$
Although we were not able to solve this equation for all $k$, we did obtain the 
general solutions for $k = 0$ and $k = -1/3$. The local and global properties of
these solutions were studied in detail, and it was found that no apparent horizons  
develop during the evolution of the fluid, although trapped regions indeed
exist. This is consistent with the general theorem obtained  previously \cite{Wang03}.

Finally we note that spacetimes with toroidal symmetry are  {\em locally} indistinguishable
from those with plane symmetry. In fact, by first unwrapping the angular coordinates $\theta$
and $\varphi$ and then extending the ranges   to $x^{A} \in (-\infty,\; \infty)$,
we    obtain spacetimes with plane symmetry. With a  different identification of coordinates,
one may obtain topologies other than  toroidal and plane geometries.

\newpage

 \end{document}